\documentclass[%
 reprint,
 amsmath,amssymb,
 aps,
]{revtex4-2}
\usepackage[colorlinks,linkcolor=blue,urlcolor=blue,citecolor=blue]{hyperref}
\usepackage{graphicx}
\usepackage{dcolumn}
\usepackage{bm}
\usepackage{subfigure}
\begin{document}

\preprint{APS/123-QED}

\title{Self-enhanced coherent harmonic amplification in seeded free-electron lasers}

\author{Hanxiang Yang\textsuperscript{1,}\textsuperscript{2}}
\author{Jiawei Yan\textsuperscript{3}}
\email{jiawei.yan@xfel.eu}
\author{Haixiao Deng\textsuperscript{4,}}%
\email{denghx@sari.ac.cn}
\affiliation{\textsuperscript{1}Shanghai Institute of Applied Physics, Chinese Academy of Sciences, Shanghai 201800, China\\
	\textsuperscript{2}University of Chinese Academy of Sciences, Beijing 100049, China\\
	\textsuperscript{3}European XFEL, Schenefeld 22869, Germany\\
	\textsuperscript{4}Shanghai Advanced Research Institute, Chinese Academy of Sciences, Shanghai 201210, China
}%

\date{\today}

\begin{abstract}
High-intensity, ultrashort, fully coherent X-ray pulses hold great potential for advancing spectroscopic techniques to unprecedented levels. Here, we propose a novel scheme for generating high-brightness and femtosecond-scale soft X-ray radiation within a seeded free-electron laser (FEL) operating at an MHz repetition rate. This scheme relies on the principles of self-modulation and superradiance. A relatively weak energy modulation of the pre-bunched electron beam is significantly amplified by the coherent radiation emitted in the self-modulator. Consequently, a coherent signal at ultra-high harmonics of the seed is achieved, and this signal is further amplified in the subsequent radiator through the fresh bunch and superradiant processes. Based on the parameters of the Shanghai soft X-ray FEL facility, three-dimensional simulations have been performed. The simulation results demonstrate that an electron beam with a laser-induced energy modulation as small as 2.3 times the slice energy spread can generate ultrashort coherent radiation pulses of around 2 GW within the water window spectral range. Moreover, the experimental results demonstrate that self-enhanced coherent energy modulation enables the production of coherent signals up to the 15th harmonic of a 266-nm seed laser. These findings indicate that the proposed scheme can facilitate the generation
of high-repetition-rate seeded FEL.
\end{abstract}

\maketitle


\section{\label{sec:1}Introduction}
X-ray free-electron lasers (XFELs) are indispensable instruments in various cutting-edge scientific fields, including materials science, energy catalysis, biomedicine, and atomic physics, owing to their ability to generate ultrashort and high-brightness radiation pulses \citep{Huang2021}. The femtosecond timescale of XFEL pulses facilitates the exploration of structural dynamics in materials, while high peak power is essential for non-linear optics \citep{Schreck2015}. In particular, soft X-ray pulses within the water window play a pivotal role in bioimaging as they can excite the core electrons of carbon and oxygen. Nevertheless, the effectiveness of most phase-related experiments utilizing single-shot pulses is limited by low collection efficiency. Therefore, the availability of ultrashort, intense, fully coherent soft X-ray pulses with high repetition rates is critical for photon-hungry experiments like resonant scattering and time-resolved pump-probe experiments \citep{Schaper2021}.

Most X-ray FEL facilities worldwide are based on the self-amplified spontaneous emission (SASE) process \citep{Kondratenko1980}, whose amplification starts from the shot noise of the electron beam. The SASE scheme can generate femtosecond radiation pulses with a peak power of tens of gigawatts but lacks temporal coherence and output stability. The self-seeding techniques \citep{Gianluca2011} are proposed to improve the longitudinal coherence of the SASE scheme but still suffer from shot-to-shot intensity fluctuations. An alternative way to enhance the coherence properties relies on the manipulation of the electron beam longitudinal phase space by externally seeding schemes, such as high-gain harmonic generation (HGHG) \citep{Yu1991}, echo-enabled harmonic generation (EEHG) \citep{Stupakov2009}, and phase-merging enhanced harmonic generation (PEHG) \citep{Deng2013}. Theoretical analyses and experiments demonstrate that seeded FELs can generate high-power and narrow-band soft X-ray radiation with small energy fluctuations.

In recent years, the generation of intense X-ray radiation pulses at the MHz-level repetition rate has been proposed to promote the application of FELs in spectroscopic experiments that require high average brightness. With the development of superconducting technology, several FEL facilities have been developed or are under construction. FLASH \citep{Ackermann2007} and European XFEL \citep{Decking2020} are operated at 1 MHz and 4.5 MHz in a burst mode, respectively. Furthermore, LCLS-II \citep{Stohr2011}, SHINE \citep{Huang2023,Liu2023}, and MariX \citep{SERAFINI2019} proposed to reach a 1 MHz repetition rate in a continue-wave mode. However, seeded FELs are difficult to operate at such high repetition rates since it is challenging for the state-of-the-art laser system to simultaneously meet sufficient energy modulation and operate at MHz repetition rates \citep{Hoeppner2015,Mecseki19}.

Various scenarios have been proposed to achieve high-repetition-rate seeded FELs \citep{Ackermann2020,Jia2021}, such as the self-modulation scheme \citep{Yan2021,yan20212,Paraskaki202100,zhao2022}, the seeded oscillator-amplifier scheme \cite{Paraskaki2021,Sun2022}, and the modulator lengthening scheme \citep{duning2011,Wang2022}. Recently, the proof-of-principle experiment on self-modulation in an HGHG setup demonstrated that a significant reduction in the peak power requirement of an external seed laser can by approximately one order of magnitude \citep{Yan2021,Yang2023}. However, due to limitations in harmonic up-conversion efficiency, the self-modulation HGHG faces challenges in generating soft X-ray radiation pulses within the water window region. Another proposed scheme, the high-brightness HGHG (HB-HGHG) \citep{Zhou2017} based on the superradiance \citep{Giannessi2005,Labat2009} and the fresh bunch technique \citep{Yu1997}, aims to generate soft X-ray pulses with durations of 20 fs and peak powers of 10 GW. However, this scheme necessitates an ultraviolet (UV) seed laser with an order of magnitude of 10 GW, posing challenges for practical implementation, especially for high-repetition-rate operations.

In this study, we propose a novel scheme to deliver ultrashort and fully coherent radiation pulses within the water window region at MHz repetition rates. Compared with the standard HGHG scheme, the proposed scheme exhibits high harmonic up-conversion efficiency within a single-stage setup and has the potential to soft X-ray pulses directly from the UV seed laser. Moreover, this scheme requires only one seed laser and is compatible with the EEHG setup, making it more feasible and straightforward to implement in the seeding beamline of existing X-ray FEL facilities.

This paper is structured as follows: the principle of the proposed scheme are described in Sec.~\ref{sec:2}. The requirements of the seed laser to implement the proposed scheme are discussed in Sec.~\ref{sec:3}. Three-dimensional simulations using the parameters of the Shanghai soft X-ray FEL facility (SXFEL), China's first X-ray FEL facility, are carried out in Sec.~\ref{sec:4}. The preliminary experimental results that verify the feasibility of the proposed scheme are shown in Sec.~\ref{sec:5}. The conclusions and outlook are finally given in Sec.~\ref{sec:6}.
\begin{figure*}
\includegraphics[width=0.95\textwidth]{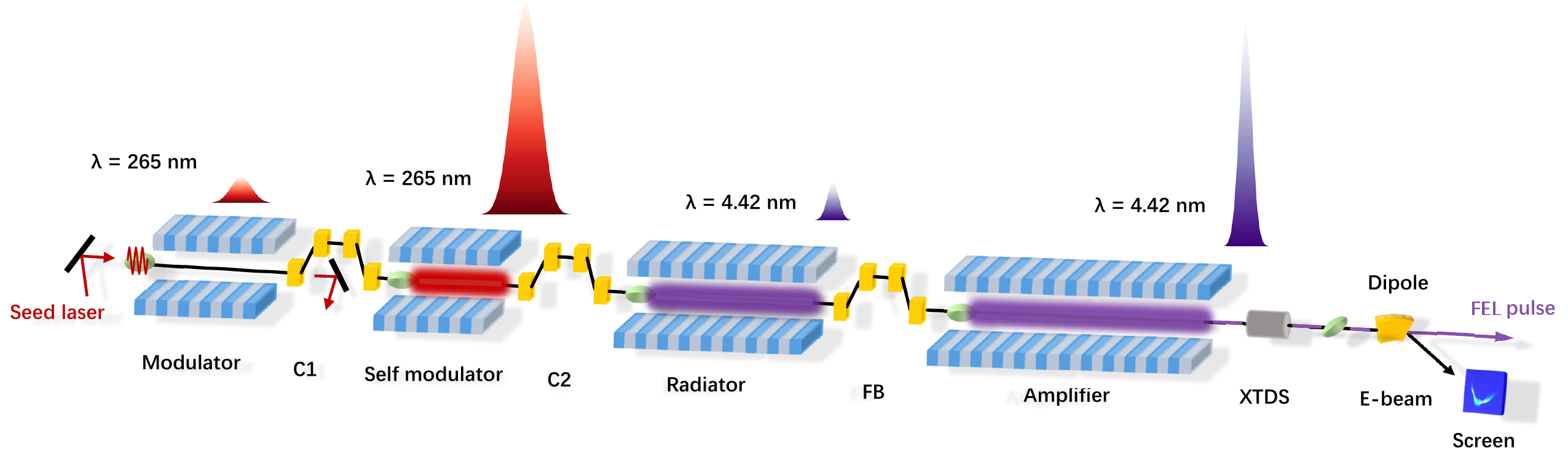}
\caption{Schematic layout of the proposed scheme. C1: the first chicane. C2: the second chicane. FB: the fresh bunch chicane. XTDS: X-band transverse deflection structure.}\label{fig:1}
\end{figure*}

\section{\label{sec:2}Principle}
In a typical seeded FEL scheme such as HGHG \citep{Yu1991}, the energy modulation induced by an external seed laser is accomplished in a modulator. Subsequently, a magnetic chicane is employed to achieve density modulation of the electron beam, which can be quantified by the bunching factor. The $n$-th harmonic bunching factor can be derived as:
\begin{equation}
    b_n=\left|J_n\left(nAB\right)\right|\exp{\left(-\frac{n^2B^2}{2}\right)}\label{eq:01}
\end{equation}
where $J_n$ is the $n$-th order Bessel function, $A = \Delta \gamma/\sigma_{\gamma}$, and $B = kR_{56}\sigma_{\gamma}/\gamma$ denotes the dimensionless energy modulation amplitude and dispersion parameter, respectively. $\gamma$ is the Lorentz factor, $\Delta \gamma$ represents the energy modulation amplitude induced by a seed laser, $\sigma_{\gamma}$ is the slice energy spread, $R_{56}$ is the dispersion strength of the chicane, and $k$ is the wavenumber of the seed laser. It is preferably for $A$ to exceed $n$ to attain a sufficient bunching factor at $n$-th harmonic. However, a dilemma arises in achieving lasing at high harmonics due to conflicting requirements. On the one hand, a large energy modulation is necessary. On the other hand, the exponential growth of FEL pulses cannot be achieved, and the peak power diminishes significantly when the relative energy spread exceeds the FEL Pierce parameter \citep{Huang2007}. To overcome this challenge, we propose a combination of the self-modulation and HB-HGHG schemes, which aims to generate fully coherent soft X-ray pulses with high repetition rates using a relatively weak energy modulation in the electron beam and a reduced seed laser power requirement.

Figure~\ref{fig:1} illustrates the schematic layout of the proposed scheme. An ultrashort UV seed laser with a relatively lower peak power is adopted to imprint a weak sinusoidal energy modulation as small as one to two times the slice energy spread into the electron beam in the first modulator. Subsequently, as the electron beam traverses a dispersive section, the energy modulation transforms into density modulation along the longitudinal beam position. The pre-bunched beam is then transferred into the additional self-modulator to significantly enhance the initial laser-induced energy modulation. This amplification leads to an ultra-large energy modulation that can be converted into an associated density modulation by another dispersive section, thereby generating extensive Fourier harmonic components of the electron beam. Following these processes, a coherent radiation signal at an ultra-high harmonic is produced by the bunched beam in the short radiator. This phenomenon can be estimated by \citep{Yu2002}
\begin{equation}
    P_{\mathrm{coh\ }}=\frac{Z_0\left(K\left[JJ\right]_1LIb_n\right)^2}{32\pi\sigma_x^2\gamma^2}\label{eq:02}
\end{equation}
where $Z_{0} = 377 \Omega$ is the vacuum impedance, $K$ is the undulator parameter, $[JJ]_{1}$ is the planar undulator Bessel factor, $L$ is the radiator length, $b_{n}$ is the $n$-th bunching factor, $I$ is the peak current, and $\sigma_{x}$ refers the transverse beam size. Coherent harmonic generation (CHG) is strongly coupled with the transverse beam size, the peak current, and the undulator length.

Notably, the output power of CHG, as described by Eq.~\ref{eq:02}, is independent of the initial energy spread. Consequently, a ultra-large energy modulation amplitude after self-modulation is acceptable. Hence, in the proposed scheme, the initial laser-induced energy modulation can be amplified as much as possible to achieve a large ultra-high harmonic bunching factor at the entrance of the radiator. This coherent signal reseeds the fresh bunch in the following amplifier segments, facilitating the further amplification of the superradiant spike until saturation is achieved. The dynamics of an isolated spike of radiation in the superradiant regime depend on appropriate slippage and harmonic up-conversion process \citep{Giannessi2005, Finetti2017, Yang2020}. The proposed scheme bears resemblance to the schematic layout of the superradiant cascade in a seeded FEL \citep{Mirian2021}.

\begin{table}
	\caption{\label{tab:table1}Main parameters of the SXFEL user facility.}
	\begin{ruledtabular}
    \begin{tabular}{lcc}
    Parameters				&Value	&Unit\\ \hline
    {\textbf{\textit{Electron beam}}}&		&\\
    Energy					&1.6		&GeV\\
    Slice energy spread 	&50     	&keV\\
    Normalized emittance  	&0.6	        &mm$\cdot$mrad\\
    Bunch charge			&500	    &pC\\
    Bunch length (FWHM)		&600    	&fs\\
    Peak current (Gaussian) &900	    &A\\
    {\textbf{\textit{Seed laser}}}&		&\\
    Wavelength                  &265		&nm\\
    Peak power  &1		&MW\\
    Pulse duration (FWHM)    	&30     	&fs\\
    Rayleigh length           	&2.96	        &m\\
    {\textbf{\textit{Modulator and self-modulator}}}&		&\\
    $K$                         &11.30		&\\
    Period                      &8		    &cm\\
    Length                      &2		&m\\
    {\textbf{\textit{Dispersion section}}}&		&\\
    $R_{56}^{1}$                         &0.776		&mm\\
    $R_{56}^{2}$                      &7.5		    &$\mu$m\\
    {\textbf{\textit{Fresh bunch section}}}&		&\\
    $R_{56}$                         &40		&$\mu$m\\
    {\textbf{\textit{Radiator}}}&		&\\
    $K$                         &2.32		&\\
    Period                     	&2.35     	&cm\\
    Length           	        &2.5	        &m\\
    {\textbf{\textit{Amplifier}}}&		&\\
    $K$                         &2.32		&\\
    Period                     	&2.35     	&cm\\
    Length           	        &7.6	        &m\\
    \end{tabular}
    \end{ruledtabular}
\end{table}

\section{\label{sec:3}Requirements of the seed laser}
Notably, a seed laser with lower peak power holds promise for enabling the proposed scheme to operate at high repetition rates. Moreover, the use of an ultrashort seed laser is expected to generate Fourier transform-limited pulses, which can benefit ultrafast non-linear spectroscopic techniques \citep{Schaper2021}. To demonstrate the feasibility of the proposed scheme and explore the optimum seed laser parameters, we consider the main parameters of the SXFEL user facility \citep{Feng22}. Currently, the baseline design of the seeding beamline of the SXFEL involves using either the two-stage cascaded HGHG or EEHG scheme to achieve a harmonic up-conversion number of nearly 60, extending to the water window spectral range. The parameters of the electron beam, undulator, and the optimized parameters of the seed laser are listed in Table~\ref{tab:table1}.

\begin{figure}
\begin{center}
\includegraphics[width=0.4\textwidth]{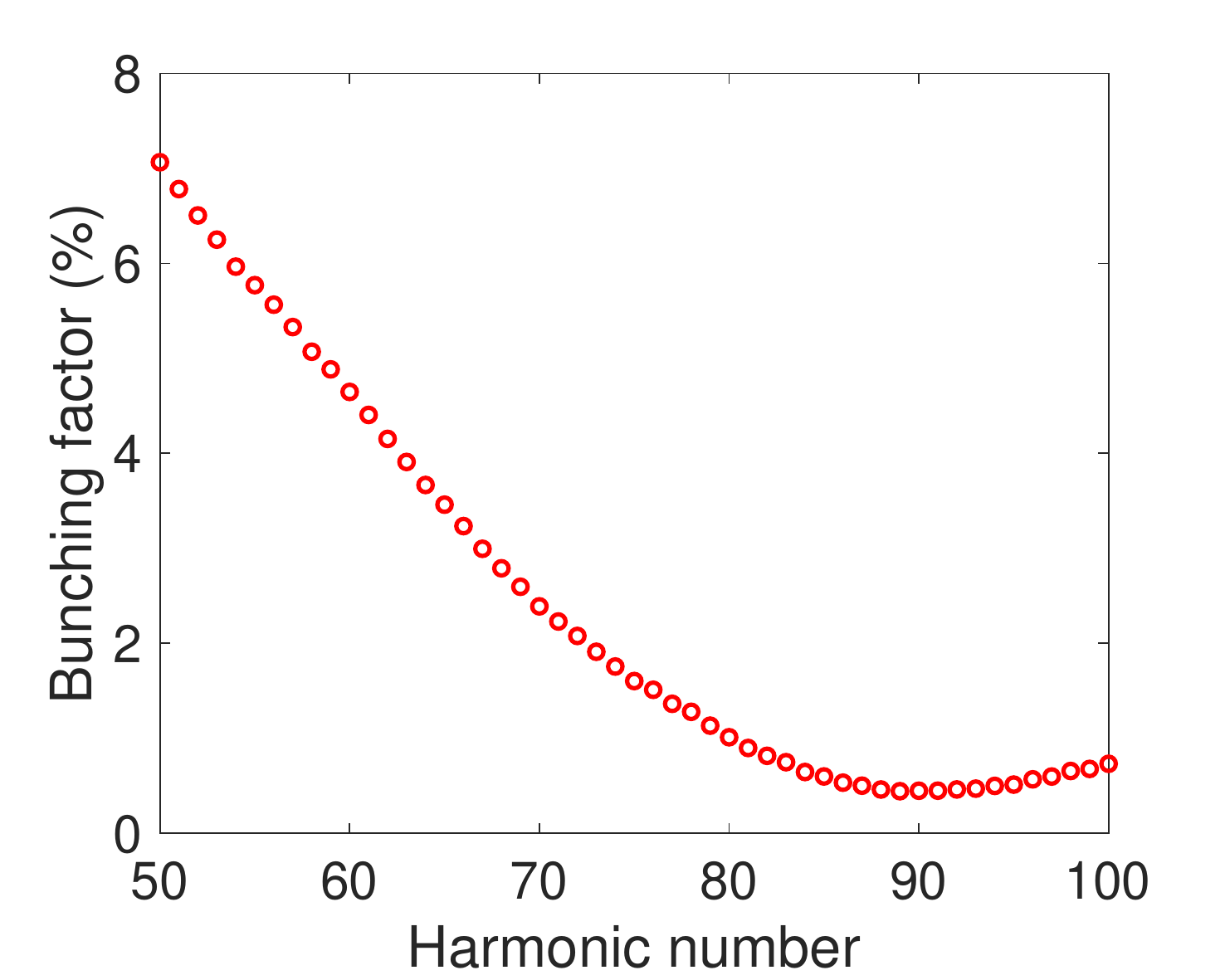}
\end{center}
\caption{The evolution of the bunching factor with harmonic numbers at the entrance of the radiator. The red dots are GENESIS \citep{REICHE1999} simulation results, and the target wavelength is the 60th harmonic of the seed laser.}\label{fig:2}
\end{figure}

In the proposed scheme, the initial energy modulation induced by a seed laser with a peak power as small as 1 MW can be significantly enhanced through the self-modulation process. As shown in Fig.~\ref{fig:2}, a bunching factor of nearly 4.6\% at the 60th harmonic of the seed laser can be achieved at the entrance of the radiator. This bunching factor is sufficient to initiate a coherent signal in the subsequent radiator. Furthermore, the observed bunching factor oscillation is due to the self-modulation that changes the energy spread distribution \citep{Ferrari2014,Wang2015}. 

In addition to achieving sufficient energy modulation, the seed laser power should be much higher than the equivalent shot-noise power \citep{Marinelli2008}. In our cases, we employ two relatively long modulators with a period number of 25 to reduce the seed laser requirement. The shot noise of the electron beam is also amplified in the self-modulator. The output power from the electron beam shot noise in the self-modulator can be estimated by \cite{Huang2007}
\begin{equation}
    P_{\mathrm{out}}\approx\frac{1}{9}P_{in}\exp{\left(z/L_g\right)}\label{eq:3}
\end{equation}
where $P_{in}$ represents the effective power of the shot noise and $L_g$ is the gain length in a SASE FEL. The one-dimensional estimate of the shot-noise power can be written as \citep{saldin1999physics}
\begin{equation}
    P_{\mathrm{in}}\simeq\frac{3\rho P_\mathrm{b}}{N_\mathrm{c}\sqrt{\pi\ln{N_\mathrm{c}}}}\label{eq:4}
\end{equation}
where $P_{b}$ is the electron beam power, $\rho$ is the FEL pierce parameter, $N_{c}$ is the number of cooperating electrons, $N_c=N_\lambda/\left(2\pi\rho\right)$, and $N_{\lambda}$ is the number of electrons in a radiation wavelength.

Utilizing the parameters listed in Table~\ref{tab:table1}, the FEL pierce parameter in the self-modulator, resonating at the seed laser wavelength of 265 nm, can be calculated to be approximately 0.01, corresponding to a gain length of 0.35 m. Employing Eq.~\ref{eq:4},  the effective shot-noise power can be estimated to be about 80 W. Furthermore, the output power originating from the shot noise in the self-modulator is calculated to be approximately 2.6 kW using Eq.~\ref{eq:3}. In the steady-state simulation, the peak power of the CHG radiation in the self-modulator is predicted to be around 1.1 GW. This power is surpasses the background signal by nearly five orders of magnitude, which is sufficient to suppress the effect of electron beam shot noise. This constraint can be satisfied when the required harmonic number is lower than 90, adhering the condition $P_{\mathrm{seed}}/P_{\mathrm{noise}} \gg n^2$ \citep{SALDIN2002169}.

\begin{figure}[h!]
\begin{center}
\includegraphics[width=0.4\textwidth]{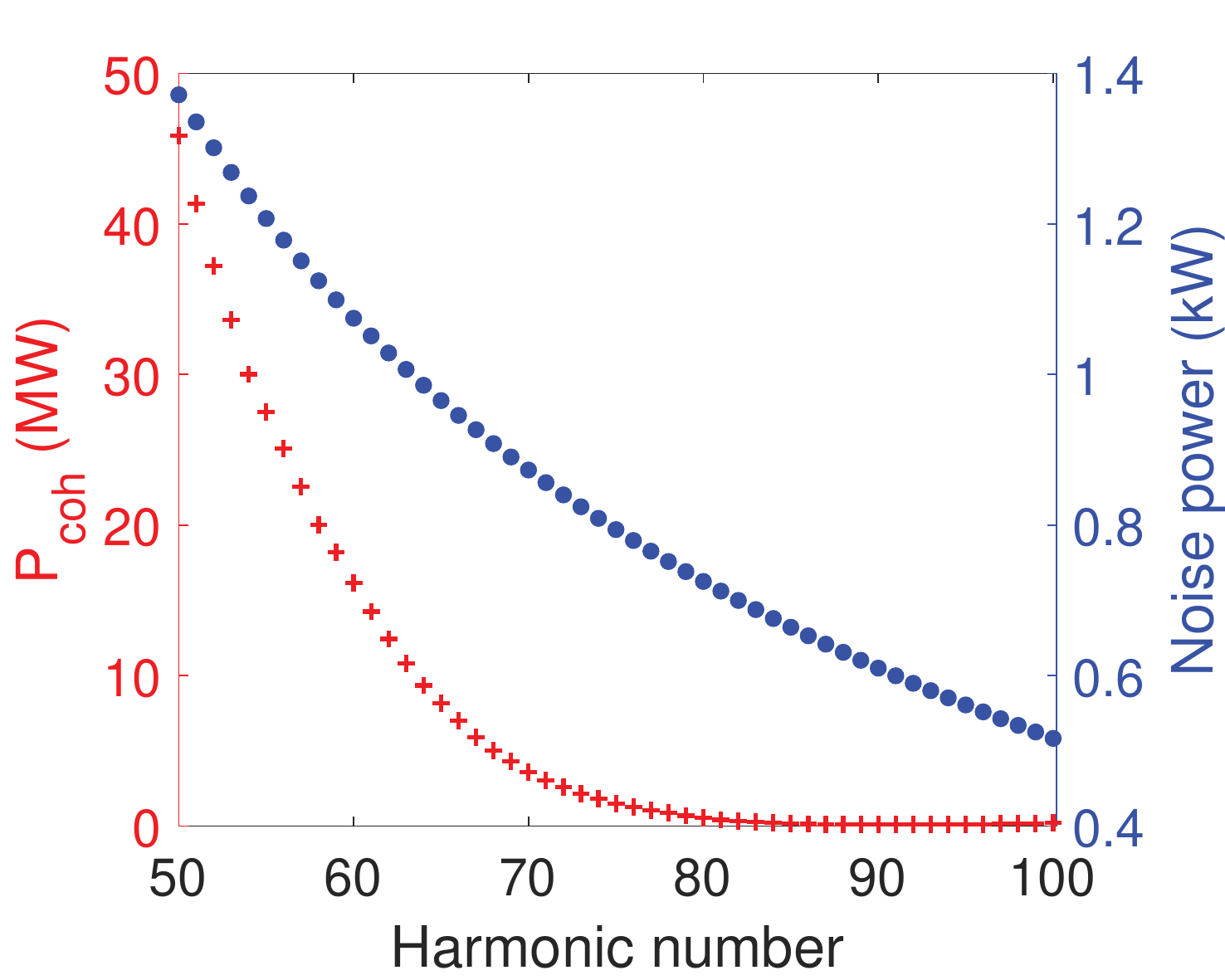}
\end{center}
\caption{Comparison of the coherent radiation power and the shot-noise power at the exit of the radiator for various harmonic numbers.}\label{fig:3}
\end{figure}

As mentioned earlier, the micro-bunched beam is sent to the radiator to generate coherent radiation at ultra-high harmonics. Upon passing through the fresh bunch chicane, a fresh part of the electron beam interacts with the coherent radiation, further enhancing the final output power in the amplifier section. Consequently, any noise introduced within the FEL finite-gain bandwidth can be amplified by the radiator and subsequent chain of amplifiers. The frequency multiplication process can exacerbate this noise, resulting in a degradation of the overall FEL performance. We conducted further investigations on the signal-to-noise ratio in the radiator, drawing an analogy to the self-modulator. The estimation of the CHG radiation generated in the radiator is determined using Eq.~\ref{eq:02}. Figure~\ref{fig:3} illustrates the calculated coherent radiation power and the corresponding noise power of the radiator at various harmonic numbers, based on the electron beam and radiator parameters listed in Table~\ref{tab:table1}. At a harmonic number of 60, the FEL pierce parameter is calculated to be about $\rho\ =\ 1.8\times{10}^{-3}$, corresponding to a gain length of 0.6 m. The peak power of the coherent signal is 16.2 MW, much larger than the noise power of 1.1 kW, satisfying the signal-to-noise ratio requirement. As shown in Fig.~\ref{fig:2}, the signal-to-noise ratio experiences a notable decrease for harmonic numbers larger than 70, primarily due to the smaller bunching factor at high harmonics. This decline is attributed to the specific optimization targeting the 60th harmonic. Nevertheless, by appropriately increasing the initial seed laser power and simultaneously enhancing the harmonic bunching factor, it becomes feasible to suppress electron beam shot noise and ultimately broaden the wavelength range of the proposed scheme.

The duration of the seed laser pulse is another crucial parameter. In a seeded FEL, the output pulse length is typically determined by the seed pulse duration, the harmonic up-conversion process, and the amplification process. For a Gaussian seed laser in the HGHG scheme, it is approximately proportional to $n^{-1/3}$ \citep{Finetti2017}. At a harmonic number of 60, a seed laser with a pulse duration of approximately 30 fs full width at half maximum (FWHM) is expected to produce nearly transform-limited FEL pulses at a level of around 10 fs. To ensure efficient energy modulation and minimize the required seed laser power, the seed pulse duration cannot be too short, as it is limited by the length of the modulator \citep{Deng2008}. Additionally, initial imperfections in the seed laser phase will be amplified during the harmonic up-conversion process, thereby degrading the quality of the output FEL \citep{Ratner2012}. However, the effects of seed laser imperfections can be mitigated by the smoothing effect resulting from the slippage in the modulator when the slippage is comparable to the seed laser duration \citep{feng2013,feng20192}. The interplay between the pulse duration of the coherent radiation and the slippage length in the self-modulator also influence the output radiation evolution. Three conditions can be distinguished: when the pulse duration is much larger than the slippage length, the standard evolution process of a seeded FEL is observed, and the steady-state model accurately describes the dynamics. However, generating ultrashort FEL pulses is challenging under these circumstances. When the pulse duration is shorter than the slippage length, the FEL in the self-modulator enters the strong superradiant regime, where the optical pulse slips ahead of the electron bunch but no longer saturates \citep{Giannessi2005, Mirian2021}. Moreover, when the pulse duration approaches the slippage length, the output FEL evolves into two sub-pulses due to strong overmodulation \citep{Labat2009}. Therefore, in our case, the period number of the first modulator and the self-modulator is set to 25, corresponding to a slippage length of around 22 fs, which mitigates the effect of the seed laser phase error. Furthermore, the 30-fs seed laser is comparable to the slippage length, enabling the FEL to operate in the superradiant regime and preserve temporal coherence.

\section{\label{sec:4}Three-dimensional simulations}
\begin{figure}
\centering
\subfigure[\label{fig:4a}]{
\includegraphics[width=0.4\textwidth]{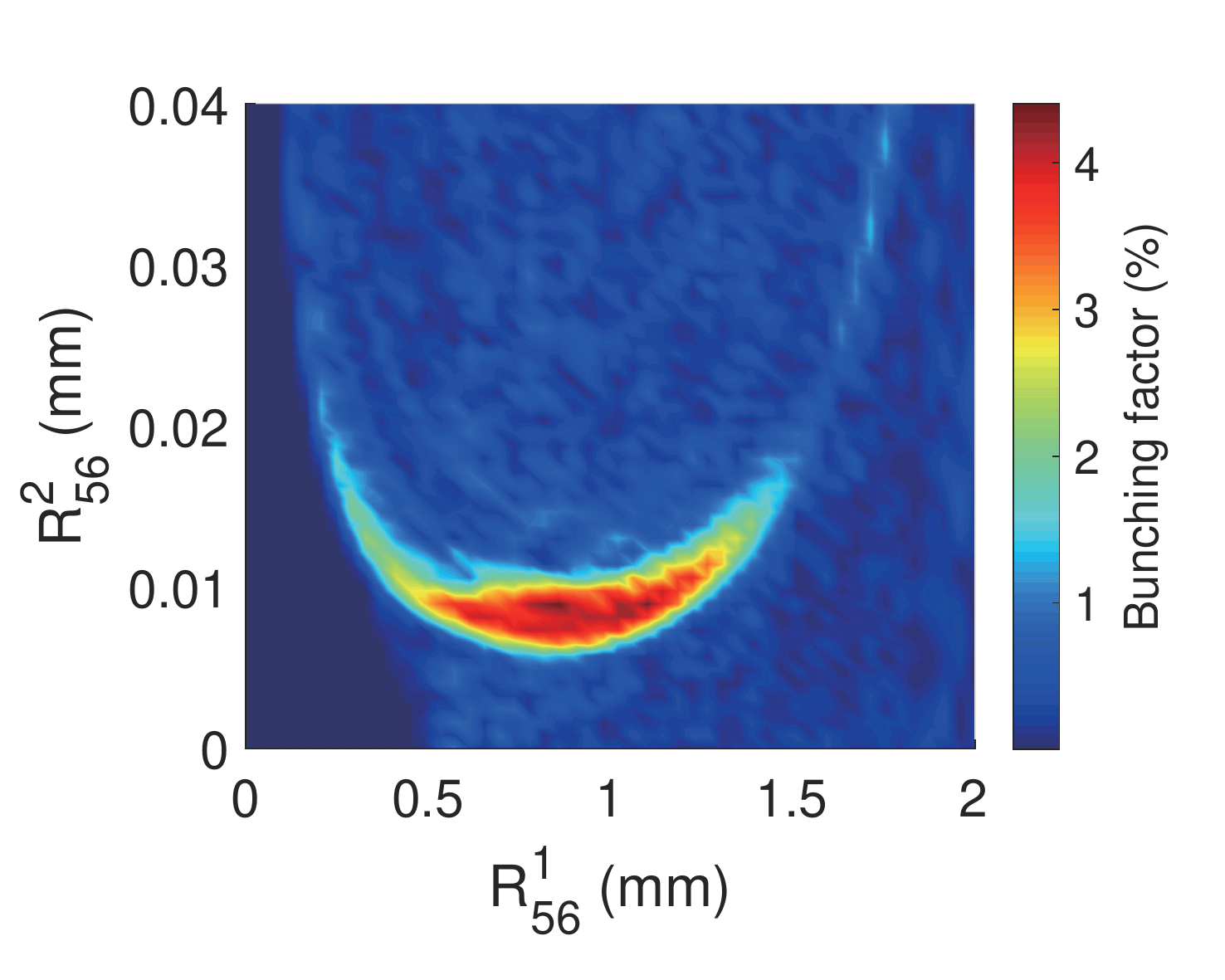}}
\subfigure[\label{fig:4b}]{
\includegraphics[width=0.4\textwidth]{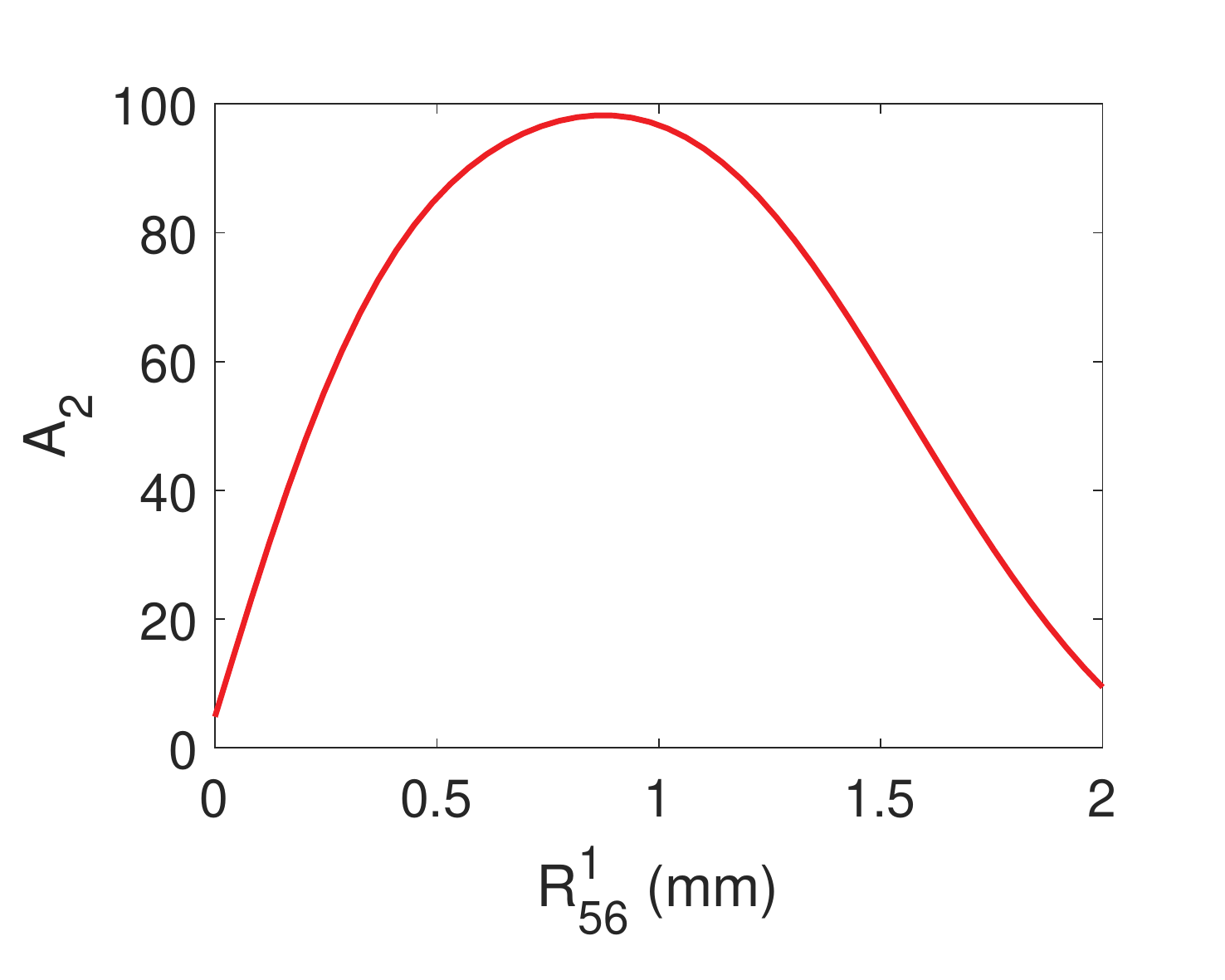}}
\caption{Optimization of the $R_{56}$ of the first and second chicanes. (a) The bunching factor at the 60th harmonic of the seed laser. (b) The enhanced energy modulation $A_2$ after self-modulator.}
\label{fig:4}
\end{figure}

To demonstrate the performance of the proposed scheme, three-dimensional time-dependent simulations are carried out with GENESIS \citep{REICHE1999} utilizing the main parameters of the SXFEL user facility, as presented in Table~\ref{tab:table1}. An ultrashort UV seed laser with a peak power of 1 MW is utilized to generate a relatively weak initial energy modulation of 115 keV in the first modulator, corresponding to 2.3 times the slice energy spread. The working points of chicane 1 (C1) and chicane (C2) are illustrated in Fig.~\ref{fig:4}. The maximum bunching factor at the 60th harmonic of the seed laser is about 4.4\%, corresponding to $R_{56}^{1}$ and $R_{56}^{2}$ values of 0.776 mm and 7.5 $\mu$m, respectively. As shown in Fig.~\ref{fig:4}, the ultra-large energy modulation amplitude $A_{2}$ reaches nearly 98.2, which can be roughly estimated by $\sqrt{2\left[\left(\sigma_{\gamma^\prime}/\sigma_\gamma\right)^2-1\right]}$.

\begin{figure*}[htbp]
    \centering
\subfigure[\label{fig:5a}]{
    \includegraphics[width=0.235\textwidth]{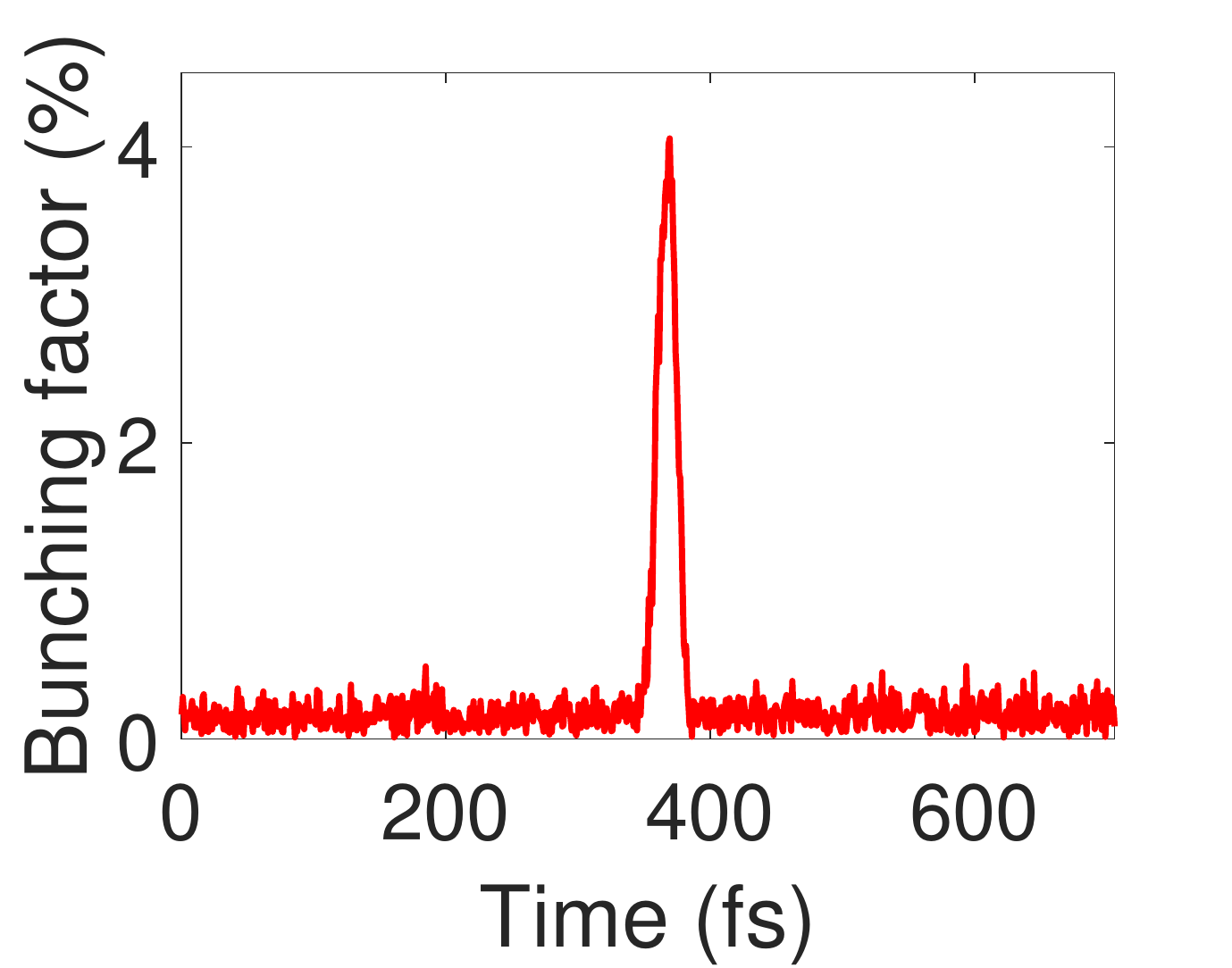}}
\subfigure[\label{fig:5b}]{
    \includegraphics[width=0.235\textwidth]{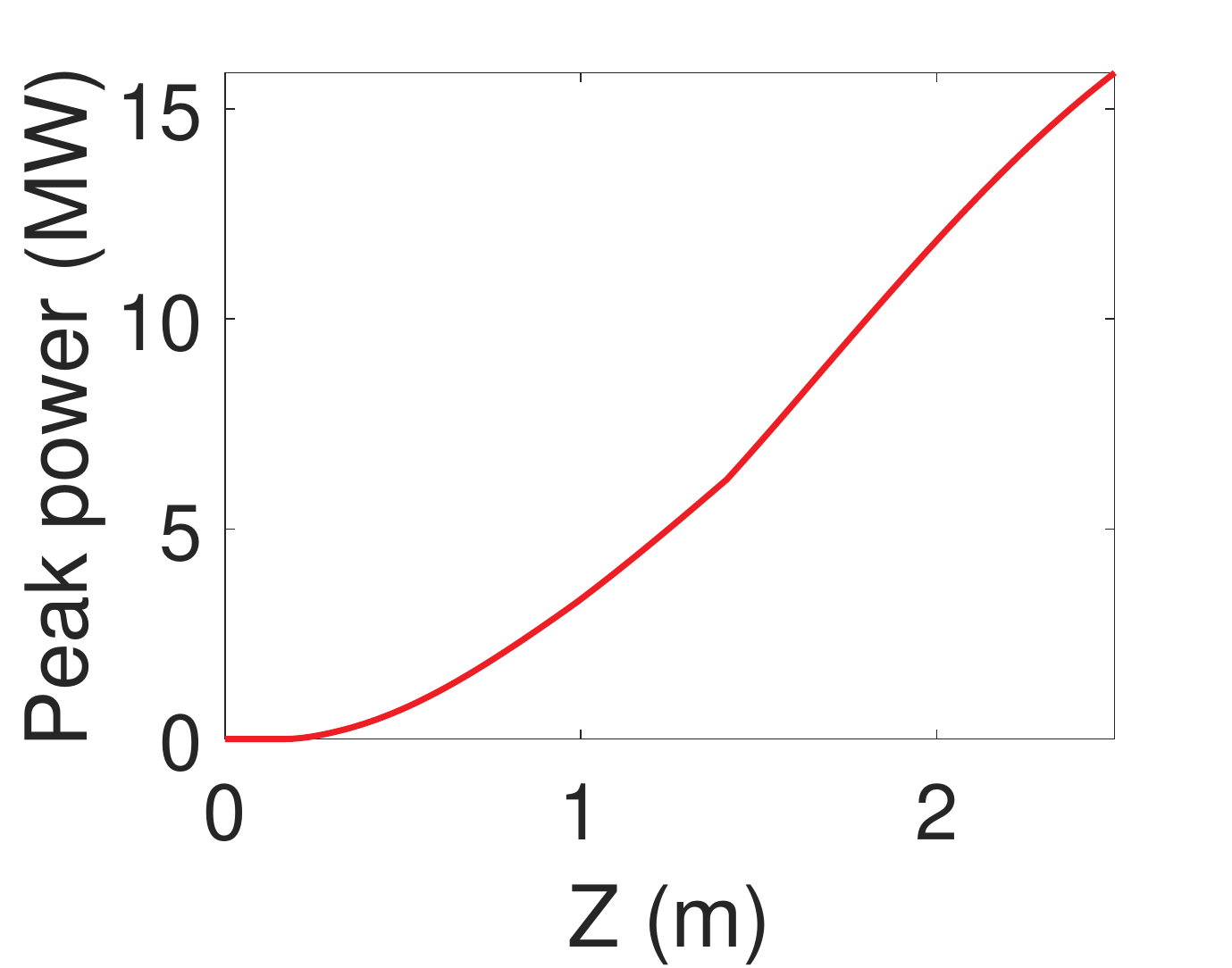}}
\subfigure[\label{fig:5c}]{
    \includegraphics[width=0.235\textwidth]{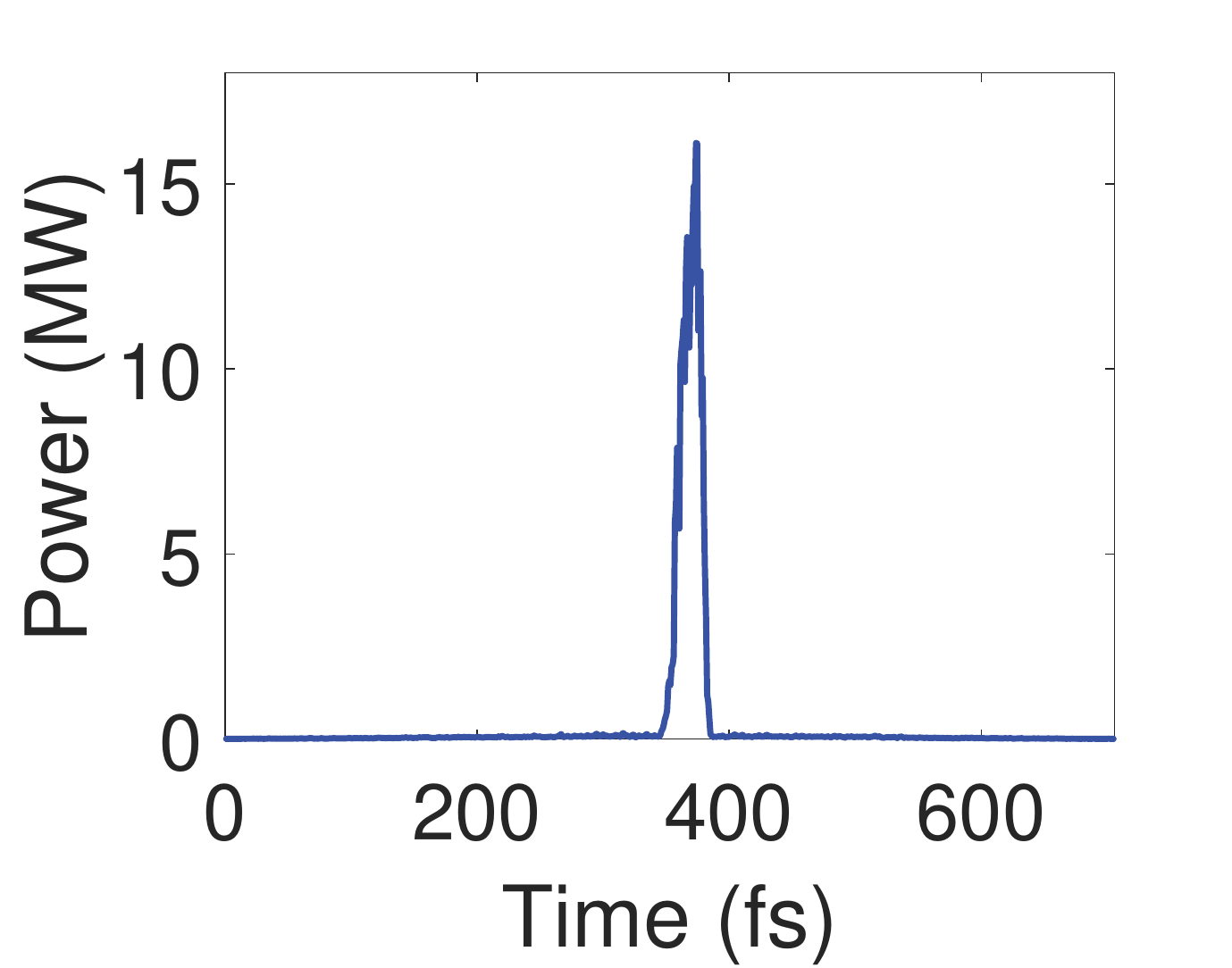}}
\quad
\subfigure[\label{fig:5d}]{
    \includegraphics[width=0.235\textwidth]{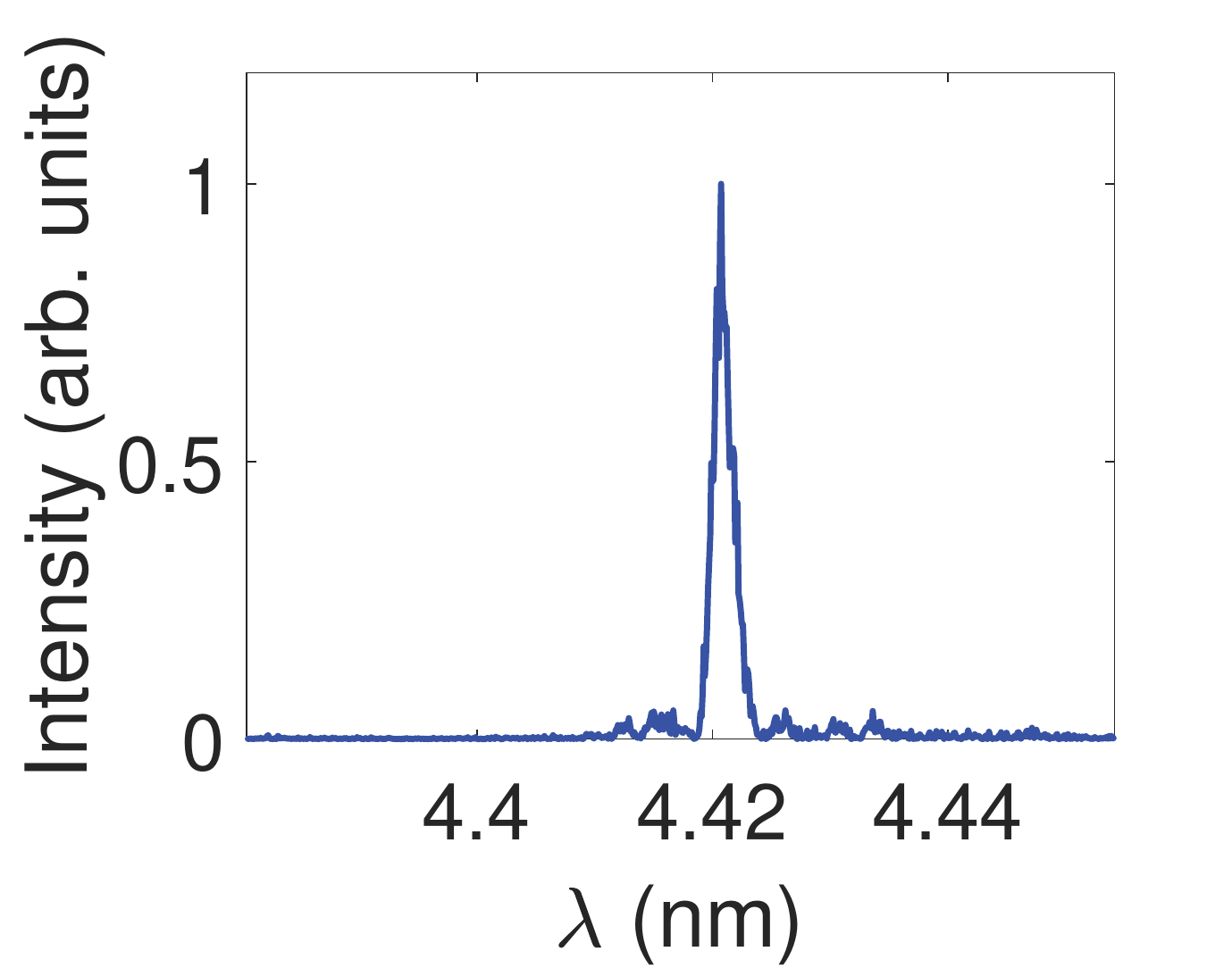}}
\quad
\caption{Simulation results of the CHG radiation in the radiator. (a): The 60th harmonic bunching factor distribution at the entrance of the radiator. (b): The gain curve of peak power evolution. (c): The power profile of CHG radiation at the exit of the radiator. (d): The spectrum of CHG radiation at the exit of the radiator.}
\label{fig:5}
\end{figure*}

\begin{figure*}
\centering
\subfigure[\label{fig:6a}]{
\includegraphics[width=0.3\textwidth]{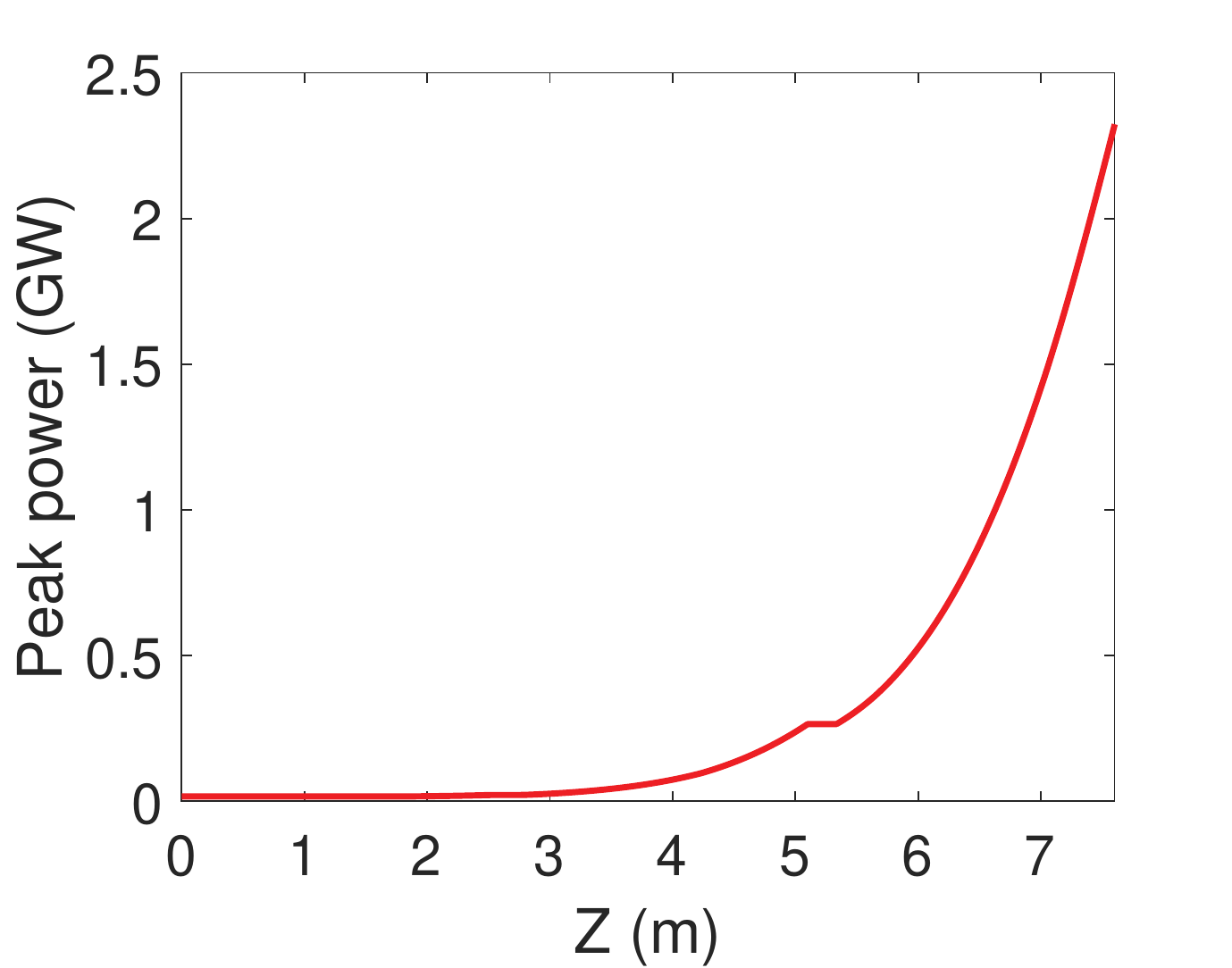}}
\subfigure[\label{fig:6b}]{
\includegraphics[width=0.3\textwidth]{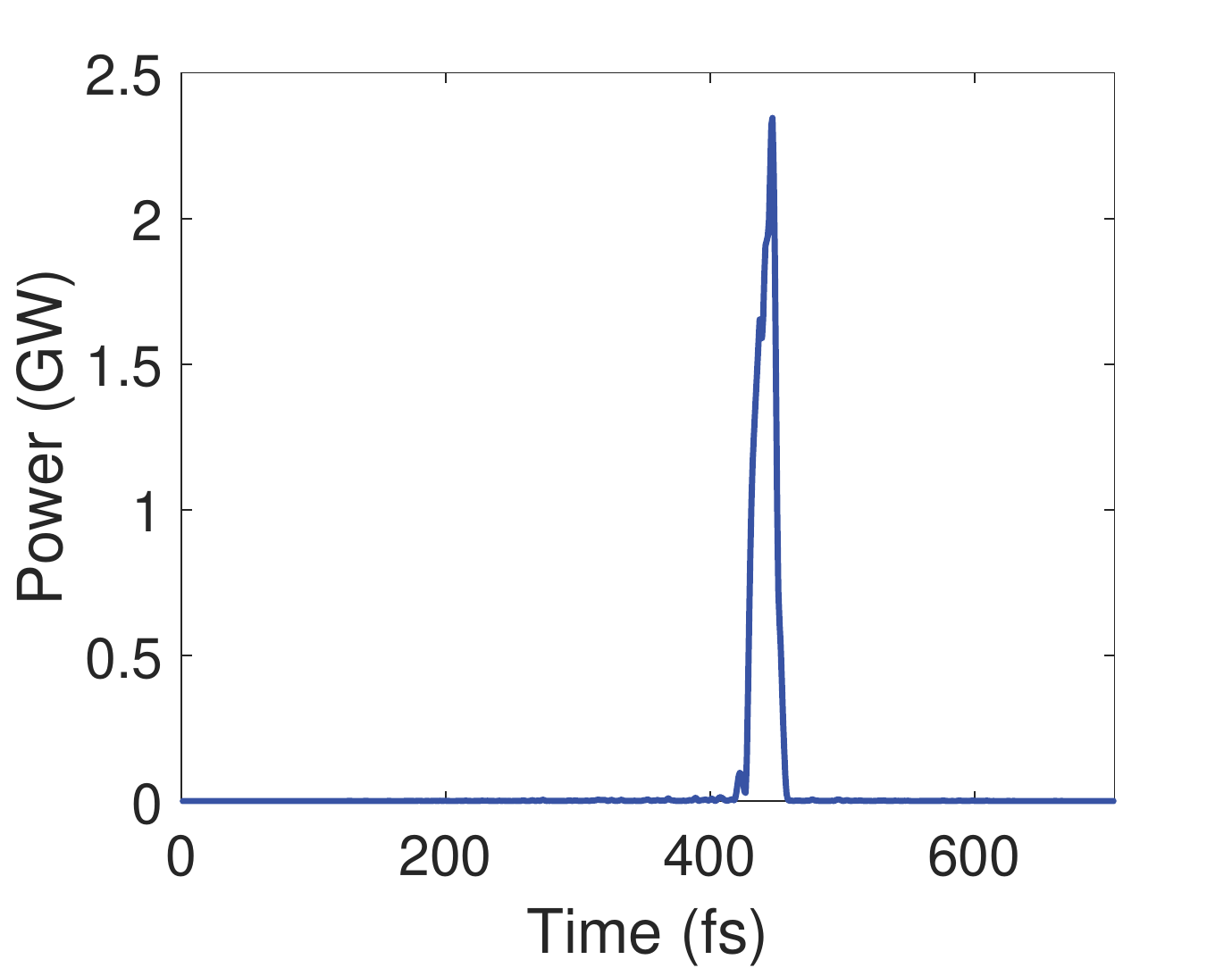}}
\subfigure[\label{fig:6c}]{
\includegraphics[width=0.3\textwidth]{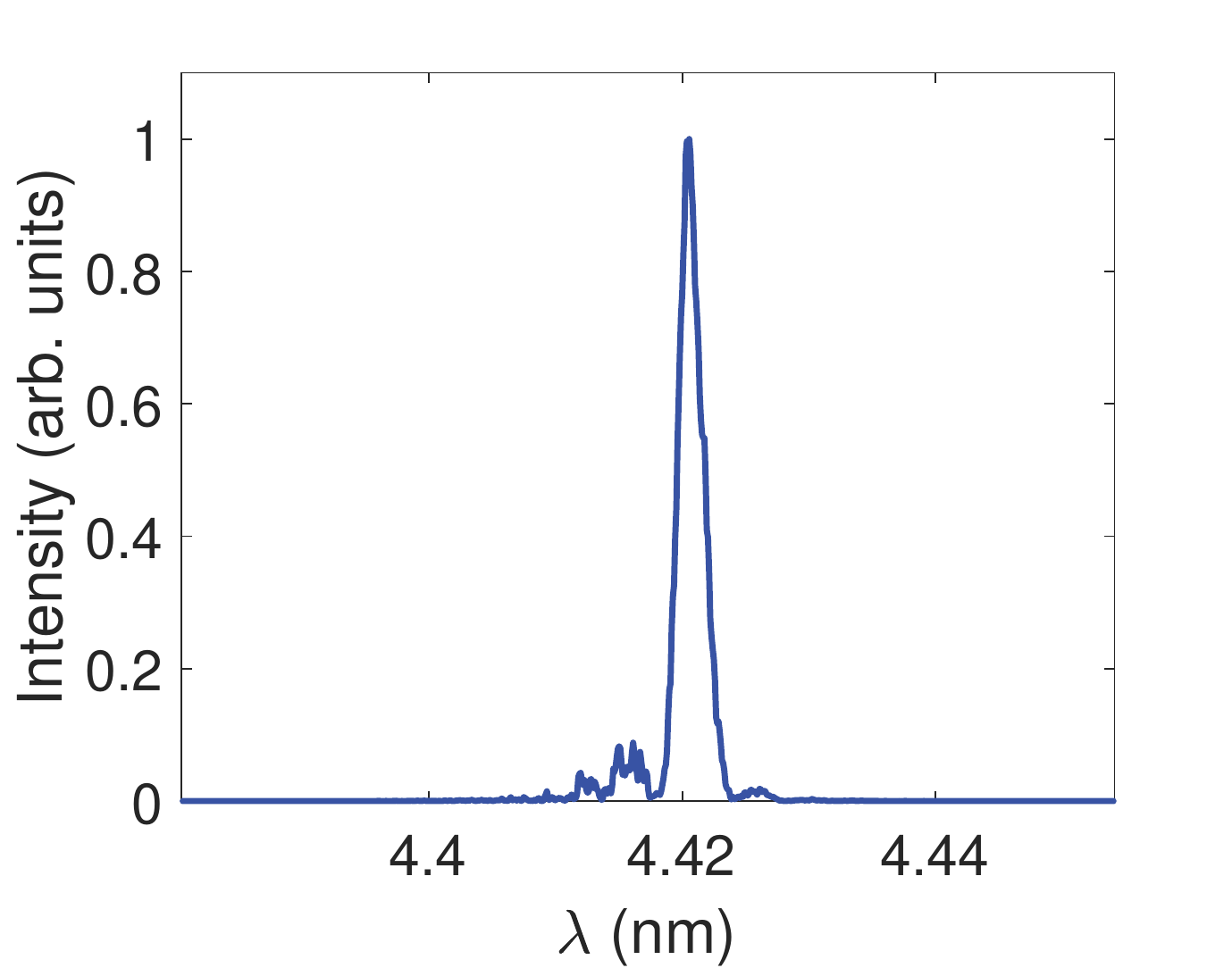}}
\caption{The final output FEL performance in the amplifier section. (a): Gain curve. (b): Power profile. (c): Spectrum.}
\label{fig:6}
\end{figure*}

Figure~\ref{fig:5a} illustrates the distribution of the 60th harmonic bunching factor along the electron beam at the entrance of the radiator. These coherent signals at ultra-high harmonic can be amplified in the following radiator, which has a length of 2.5 m. The evolution of peak power, as shown in Fig.~\ref{fig:5b}, exhibits quadratic growth \citep{Giannessi2005, Yang2020}, meeting the desired criteria. The characteristics of the CHG radiation are shown in Fig.~\ref{fig:5c} and Fig.~\ref{fig:5d}, where the output power of 16.1 MW closely aligns with the calculated results. It is observed that the pulse duration of the CHG radiation is 17.7 fs (FWHM), and the pulse leading edge exhibits a high-power spike, where the bunching is higher and the pulse slips over the fresh part of the electron beam. Furthermore, it is essential to note that the fresh bunch chicane must provide sufficient time delay and smear out the microbunching formed in the previous undulator. To suppress noise arising from the unmodulated part of the electron beam, the $R_{56}$ value of the fresh bunch chicane is set to 40 $\mu$m, corresponding to a time delay of approximately 67 fs. Subsequently, this coherent signal acting as a seed, continues to exchange energy with the fresh electrons, eventually producing an intense (2.4 GW) and short (16.8 fs FWHM) superradiant spike at the pulse leading edge, as shown in Fig.~\ref{fig:6}. Tapered undulators are employed in the amplifier section to enhanced the output power of the soft X-ray pulses. By observing the spectral sidebands in the spectrum and discerning two spikes in the tail structure of the pulse from Fig.~\ref{fig:6}, we can identify the effects of saturation and temporal splitting of the pulse caused by interference between the tail and the main peak \cite{Labat2009, Mirian2021, Yang2020}. Due to slippage in the amplifier, the bunched electron beam has a phase mismatch of approximately $\pi$ with the field of the main spike at the front leading edge of the pulse, thus leading to a second spike. This process is repeated in the portion of the electron beam modulated by the seed laser, and the amplitude of the spike is significantly reduced due to the large induced energy spread. These secondary pulses form the pedestal of the pulse, thereby generating spectral sidebands. A tailored taper can compensate for the degradation of the FEL pulses due to the phase mismatch, producing X-ray pulses approaching the Fourier transform limit. Along the following amplifier section, at a distance of 7.6 m, the rms bandwidth is 0.053\%, corresponding to the time-bandwidth product (TBP) of 1.41. Thus, the seeded XFEL in the proposed scheme exhibits nearly full coherence.

\begin{figure}
\centering
\subfigure[\label{fig:7a}]{
\includegraphics[width=0.4\textwidth]{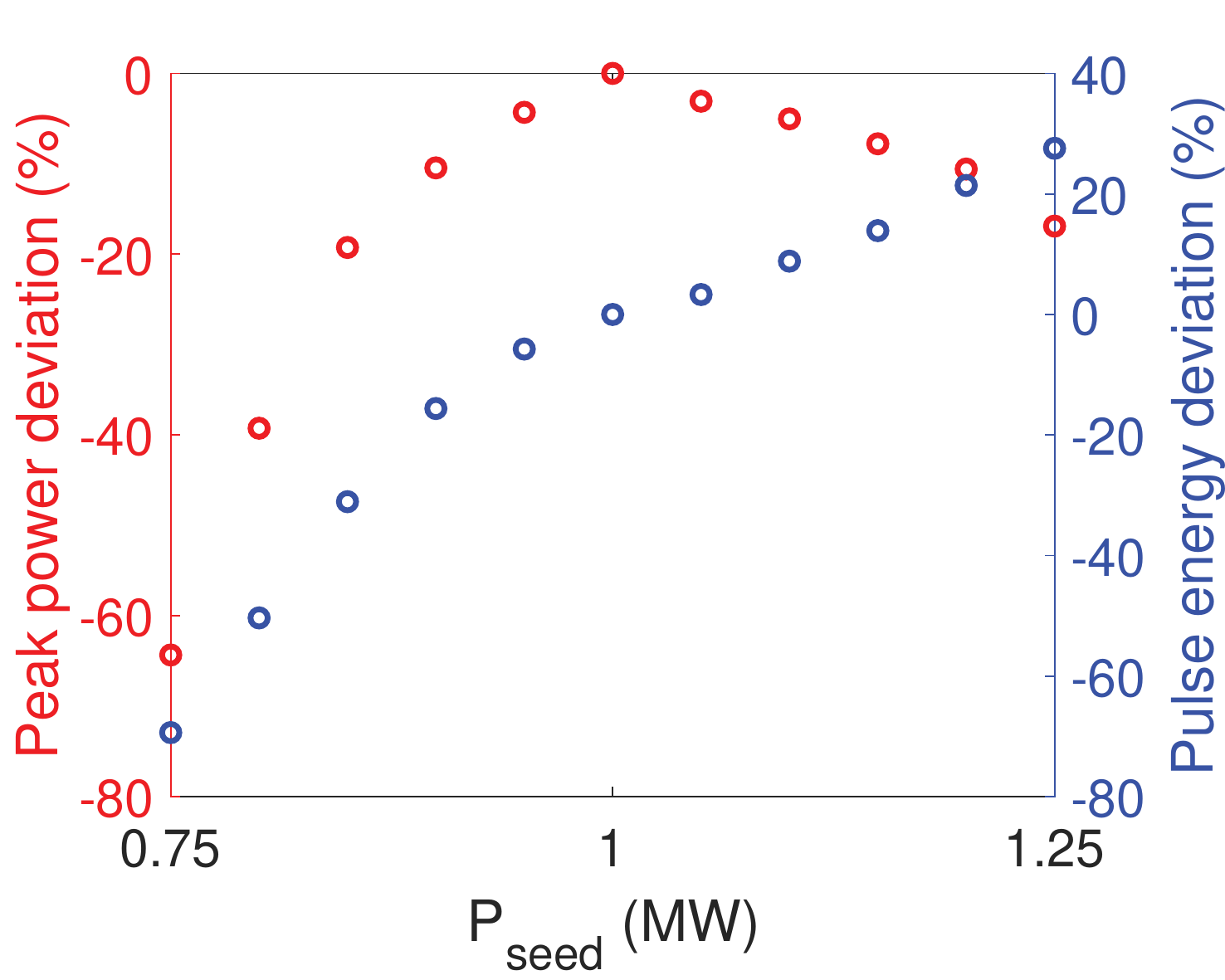}}
\subfigure[\label{fig:7b}]{
\includegraphics[width=0.4\textwidth]{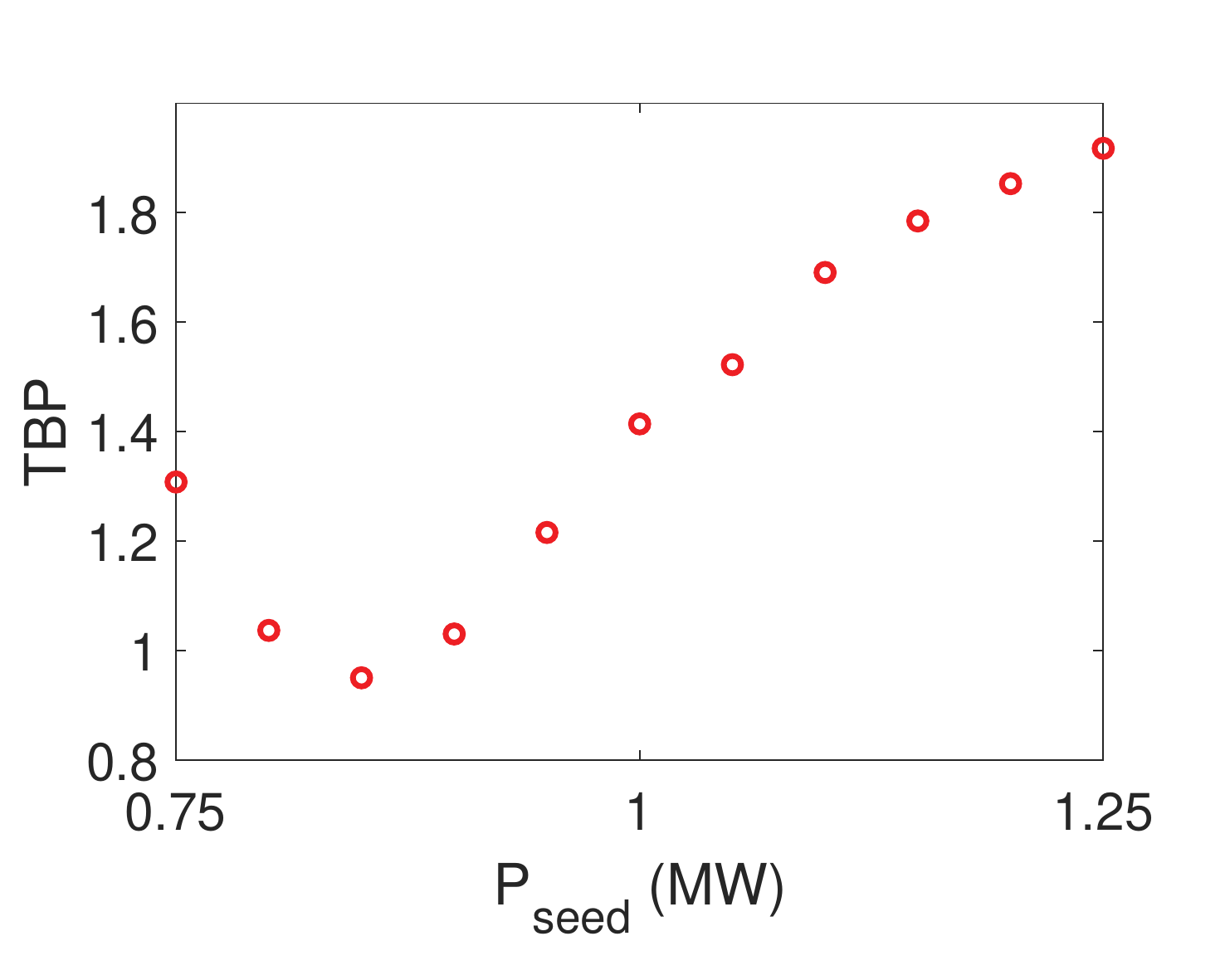}}
\caption{Sensitivity of the output FEL performance versus the seed laser power. The nominal seed laser power is 1 MW. (a): The deviation of the peak power and the pulse energy. (b): The time-bandwidth product calculated from the scaled FWHM pulse duration and rms bandwidth.}
\label{fig:7}
\end{figure}

The stability of the proposed scheme has also been investigated. Figure~\ref{fig:7} illustrates the output properties of the FEL as a function the seed laser power. We conducted a scan of the seed laser power ranging from 0.75 MW to 1.25 MW while maintaining other parameters constant, including the $R_{56}$ values of C1 and C2, the optimal tapering (for the nominal case of 1 MW), and the total length of the amplifier section. As depicted in Fig.~\ref{fig:7a}, both the peak power and the pulse energy of the FEL exhibits a significant decrease when the seed laser power falls below 1 MW. However, at a seed laser power of 0.85 MW, the TBP reaches a minimum value of 0.95. The reason is that the $R_{56}$ is not the optimal value, resulting in a dominant main peak with a large proportion and a short pulse duration. Similarly, the TBP of the output FEL pulses as a function of the seed laser power, as shown in Fig.~\ref{fig:7b}. In addition, it is noteworthy that for the seed laser powers much larger than 1 MW, the FEL spectra exhibits sidebands, and the FEL pulse energy increases due to the presence of the pulse tail and the growth of noise in the finite-gain bandwidth. The evolution of the superradiant spike is sensitive to the $R_{56}$ value of C2, owing to the ultra-large energy modulation. Therefore, by slightly over-modulating the longitudinal phase space of the electron beam by through an appropriate increase in the $R_{56}$ value of C2, a shorter and intense superradiant spike can be generated.

Given the relatively low seed laser power employed in the proposed scheme, the SASE background inevitably reduces the signal-to-noise ratio. Moreover, the phase mismatch of the bunched electron beam and the radiation field adversely affect the quality of the superradiant spike. To assess the effect of shot noise, we conducted 60 three-dimensional simulations, each with different shot noises. As anticipated, the simulations show average values of 36.1 $\mu$J and 16.3 fs for pulse energy and pulse duration, respectively, corresponding to rms pulse energy jitter and rms pulse duration jitter of 4.7\% and 4.8\%. The shot-to-shot pulse fluctuation primarily arises from the mismatch between the FEL power and the optimal taper setting. Furthermore, the average bandwidth of the spectra obtained from 60 simulations is 0.072\%, corresponding to an rms bandwidth jitter of 12.4\%. The spectral bandwidth jitter in this scheme may stem from the phase mismatch between the shot noise in the electron beam and the radiation field, as well as the temporal splitting of the pulse. In addition, different settings of the undulator taper and the  frequency-pulling effect of the superradiant pulse \citep{Mirian2021} may introduce the bandwidth jitter in the final FEL spectra. However, the implementation of optical elements such as gratings or monochromators can mitigate the FEL bandwidth jitter in time-resolved spectroscopy experiments.

\begin{figure}[h!]
\centering
\subfigure[\label{fig:8a}]{
\includegraphics[width=0.4\textwidth]{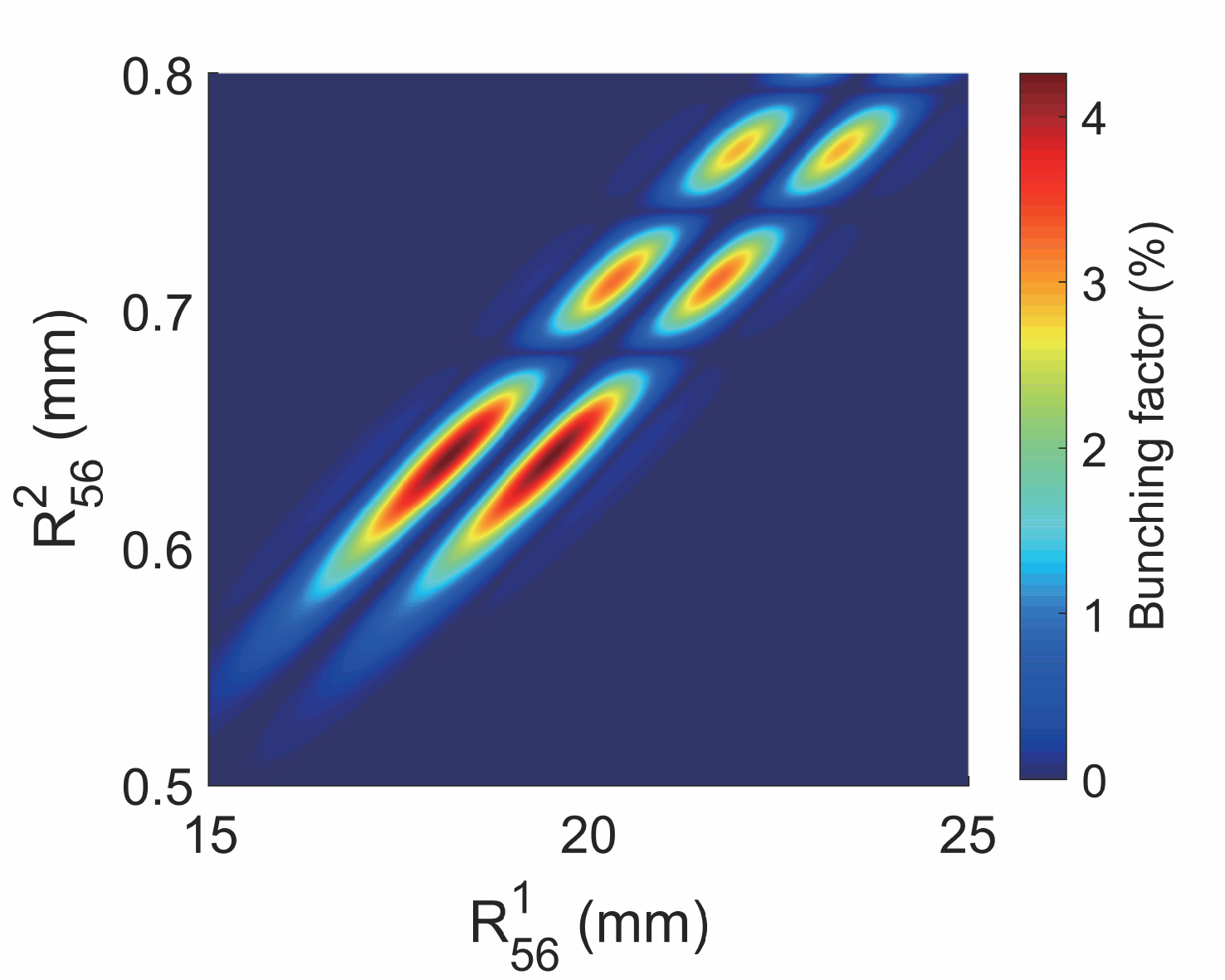}}
\subfigure[\label{fig:8b}]{
\includegraphics[width=0.36\textwidth]{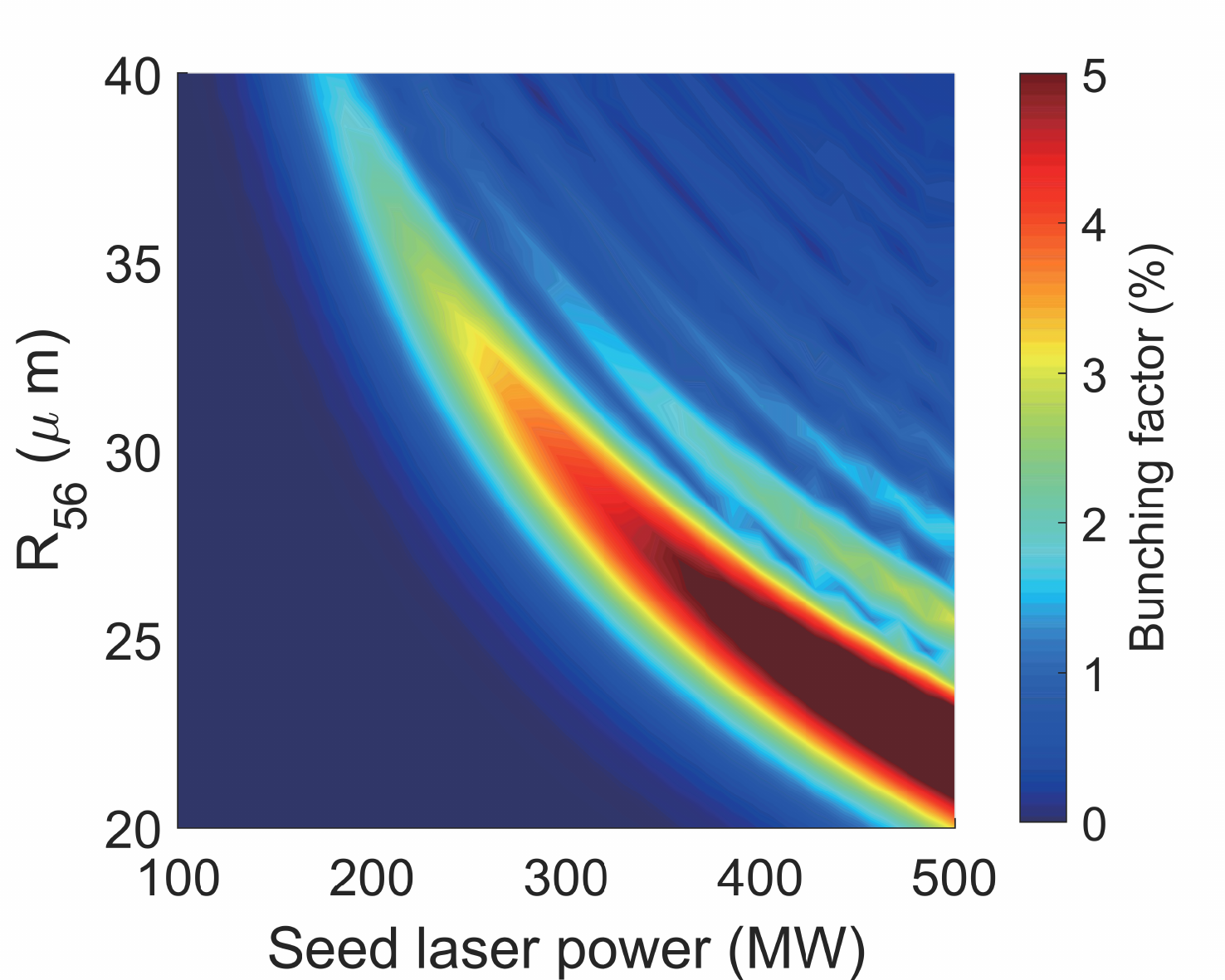}}
\caption{(a): The bunching factor at the 60th harmonic for the EEHG scheme. $A_1 = A_2 = 2.3$. (b): The bunching factor at the 60th harmonic for the HB-HGHG scheme.}\label{fig:8}
\end{figure}

To assess the advantages of the proposed scheme, we conducted a comparison with EEHG and HB-HGHG schemes, respectively. Figure~\ref{fig:8a} presents the calculation results of the 60th harmonic bunching factor for EEHG in the high-order mode $n = -2$, considering a laser-induced energy modulation amplitude of 2.3 for both modulators \citep{Zhou2016}. To order to achieve the same bunching factor of 4.4\% as the the proposed scheme, EEHG requires a significantly larger $R_{56}$ value of 17.8 mm in the first chicane, which is one order of magnitude larger than that of the proposed scheme. The chicane with a considerable dispersive strength has a non-linear effect on the electron beam leading to a degradation of the output FEL \citep{Hemsing2017}. It is important to emphasize that the layout of the proposed scheme is compatible with that of EEHG. Furthermore, the proposed scheme has the advantage of requiring only one seed laser, alleviating the burden on a high-repetition-rate seed laser system. The presence of a chicane with substantial dispersive strength in EEHG can lead to non-linear effects on the electron beam, resulting in a degradation of the output FEL quality \citep{Hemsing2017}. Additionally, we compared the proposed scheme with HB-HGHG. Figure~\ref{fig:8b} illustrates the simulated 60th harmonic bunching factor for HB-HGHG, utilizing the same modulator parameters listed in Table~\ref{tab:table1}. To achieve an equivalent bunching factor as the proposed scheme with a seed laser power of 1 MW, HB-HGHG necessitates a seed power of approximately 300 MW, which is two orders of magnitude higher than that required by the proposed scheme.

\section{\label{sec:5}Experiment}
We designed and conducted an experiment at the SXFEL test facility (SXFEL-TF) \citep{Feng22} to verify the feasibility of the proposed scheme for achieving coherent signals at high harmonics. SXFEL-TF was designed to generate fully coherent soft X-ray pulses by cascaded EEHG-HGHG scheme. We considered a modulator of the first stage EEHG with a length of 1.5 m and period of 80 mm as the first modulator and a modulator of the second stage HGHG with a length of 3 m and period of 55 mm as the self-modulator. The fresh bunch chicane was used as the first chicane, and the second stage's dispersion section was regarded as the second chicane.
\begin{figure}
\begin{center}
\includegraphics[width=0.45\textwidth]{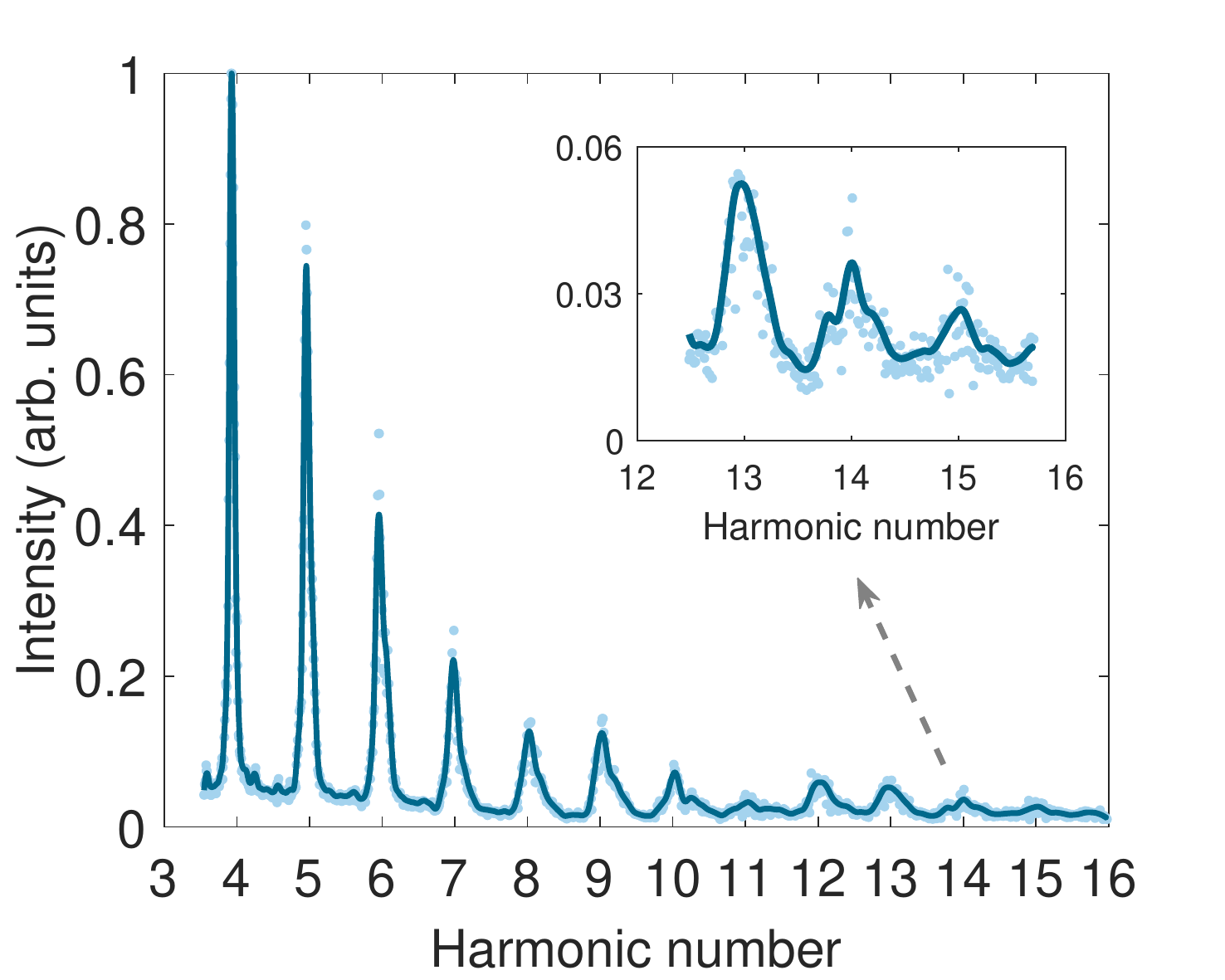}
\end{center}
\caption{Measured coherent radiation intensities at various harmonic numbers. The curve is the envelope obtained by smoothing the measurement data.}\label{fig:9}
\end{figure}

In the experiment, the electron beam has an energy of 780 MeV, a bunch charge of 550 pC, a peak current of 600 A, a normalized emittance of 1.5 mm$\cdot$mrad, a transverse beam size of 300 $\mu$m, and a slice energy spread of 40 keV, which had been measured in a previous experiment \citep{Yan2021}. A seed laser with a wavelength of 266 nm and a pulse duration of 160 fs (FWHM) was employed to interact with the electron beam in the first modulator. The $R_{56}$ value of the first chicane was set as 0.24 mm. To determine the energy modulation amplitude $A_1$, the self-modulator was removed. In the subsequent radiator, a 3 m long undulator with a period of 40 mm was utilized to detect coherent signals, specifically focusing on the resonance of the 4th to 11th harmonics of the seed laser. In this HGHG-like scenario, coherent radiation signals above the 6th harmonic were observable due to the utilization of a high-power seed laser. To ensure a weak initial energy modulation, the intensity of the seed laser was continuously attenuated until the signal at the 4th harmonic was no longer undetectable, indicating an initial energy modulation amplitude below four times that of the slice energy spread. Following that, the self-modulator was resonated at the fundamental wavelength and the focus and orbit of the electron beam were optimized to explore the self-modulation's potential for generating higher harmonics. The undulator segment gap was continuously varied from 9.2 mm to 21 mm, covering the 4th to 15th harmonics of the seed laser. Simultaneously, the $R_{56}$ value of the second chicane was optimized to 0.041 mm to enhance the coherent signals above the 12th harmonic. Figure~\ref{fig:9} presents the intensity of coherent radiation at various harmonic numbers. The coherent energy modulation at the fundamental wavelength was prominently enhanced by self-modulation, where coherent signals at high harmonics can be observed.

Since the amplification of the coherent energy modulation is not directly introduced by the external seed laser, the coherent radiation-based method \citep{Feng2011} is challenging to measure the energy modulation amplitude at the exit of the self-modulator. However, we can assume that the enhancement of coherent energy modulation is equivalent to an intense seed laser modulating an electron beam, similar to the standard HGHG. Consequently, the energy modulation amplitude $A_2$ after the self-modulator can be roughly estimated using the relationship between optimal dispersion strength and energy modulation amplitude. The coherent energy modulation amplitude was estimated to be about 925 keV corresponding to $A_2=23.1$. Since setting $R_{56}$ to 0.041 mm is not optimized for a specific wavelength, the value of $A_2$ is slightly different, and we averaged them to obtain a result of 23.3. These preliminary experimental results indicate that the proposed scheme can generate coherent signals at higher harmonics with a weak initial energy modulation. The energy modulation is enhanced by a factor of at least 6, corresponding to a nearly thirty-six-fold reduction in the peak power requirement of the seed laser. Moreover, by more careful optimization of the beam orbit and focusing, ultra-high harmonic bunching signals can be achieved and are expected to realize short-wavelength seeded FELs.

\section{\label{sec:6}Conclusions and outlook}
A novel scheme is proposed in this paper for generating soft X-ray pulses with high repetition rates, high brightness, and full coherence. Theoretical analysis and numerical simulations of the proposed scheme demonstrate that an electron beam exhibiting laser-induced energy modulation 2.3 times energy spread can achieve ultra-large energy modulation and ultra-high harmonic up-conversion efficiency. Experimental results 
demonstrate that self-enhanced coherent energy modulation enables the production of coherent radiation signals up to the 15th harmonic.

In comparison to the EEHG scheme, the proposed scheme necessitates at least one order of magnitude less dispersion strength and utilizes only one seed laser. Furthermore, in contrast to the HB-HGHG scheme, the proposed scheme reduces the peak power requirement of the seed laser by two orders of magnitude. In addition, the output power of the proposed scheme can be increased by incorporating amplifier modules and a flat-top current profile. By tuning the dispersion strength of the second chicane or incorporating the superradiant harmonic cascade, there is potential to generate few-femtosecond FEL pulses at shorter wavelengths. More importantly, the proposed scheme operates on simple principles, making it easy to implement and compatible with existing seeded FEL facilities. With the advancement of superconducting technology, MHz-repetition rate electron beams have become attainable. The scheme proposed in this paper is expected to overcome the limitations of current seed laser systems and offer new insights for the investigation of high-repetition-rate seeded FELs.

\begin{acknowledgments}
The authors would like to thank Nanshun Huang and Weijie Fan for their helpful discussions and comments. This work was supported by the CAS Project for Young Scientists in Basic Research (YSBR-042), the National Natural Science Foundation of China (12125508, 11935020), Program of Shanghai Academic/Technology Research Leader (21XD1404100), and Shanghai Pilot Program for Basic Research - Chinese Academy of Sciences, Shanghai Branch (JCYJ-SHFY-2021-010).
\end{acknowledgments}


\bibliography{reference}

\end{document}